\RequirePackage{fix-cm}

\documentclass{svjour3}                     
\smartqed  
\usepackage{overpic}
\usepackage{graphicx}
\usepackage{hyperref}
\usepackage{inputenc}
\journalname{Few-Body systems}
\begin{document}

\title{One way to verify the molecular picture of exotic hadrons --from $DK$ to $DDK/D\bar{D}^{(*)}K$} \thanks{This work is partly supported by the National Natural Science Foundation of China under Grants Nos.11735003, 11975041, and 11961141004, and the fundamental Research Funds for the Central Universities.
}

\titlerunning{$DDK$ and $D\bar{D}^{(*)}K$ bound states}        

\author{Tian-Wei Wu         \and
        Ming-Zhu Liu        \and
        Li-Sheng Geng 
}


\institute{Tian-Wei Wu \at
              School of Physics, Beihang University, Beijing 102206, China \\
 \and
Ming-Zhu Liu \at
              School of Physics, Beihang University, Beijing 102206, China \\
           \and
           Li-Sheng Geng \at
         School of Physics \& Beijing Key Laboratory of Advanced Nuclear Materials and Physics, \\
          Beihang  University, Beijing 102206, China\\
School of Physics and Microelectronics,\\
Zhengzhou University, Zhengzhou, Henan 450001, China\\
          \email{lisheng.geng@buaa.edu.cn}
}

\date{Received: date / Accepted: date}

\maketitle

\begin{abstract}
Starting from 2003, a large number of the so-called exotic hadrons, such as $X(3872)$ and
$D_{s0}^*(2317)$, were discovered experimentally. Since then, understanding the nature of these states has been a central issue both theoretically and experimentally. As many of these states are located close to two hadron thresholds, they are believed to be molecular states or at least contain large molecular components. We argue that if they are indeed molecular states, in the way that the deuteron is a bound state of proton and neutron, then molecular states of three or more hadrons are likely, in the sense that atomic nuclei are bound states of nucleons. Following this conjecture, we study the likely existence of $DDK$, $D\bar{D}K$, and $D\bar{D}^{*}K$  molecular states. We show that within the  theoretical uncertainties of the two-body interactions deduced, they most likely exist. Furthermore, we predict their strong decays to help guide future experimental searches. In addition, we show that the same approach can indeed reproduce some of the known three-body systems from the two-body inputs, such as the deuteron-triton and the $\Lambda(1405)$-$\bar{K}NN$ systems.

\end{abstract}

\section{Introduction}
\label{intro}
Up to 1932, our visible universe looked much simpler. There exist only three matter particles, electron, proton, and neutron. By combing different numbers of protons and neutrons, one can obtain all the stable atomic nuclei. The electromagnetic interaction then binds nuclei and electrons together to form atoms. Atoms can form molecules via residual electromagnetic interactions. From atoms and molecules, the whole visible universe is built up. Surprisingly,in the following years, three particles (leptons) were discovered, $e^+$ and $\mu^\pm$.  Starting from 1947,  more particles (hadrons) were discovered, including $\pi^
\pm$,  $K^\pm$, $\Lambda$, $\Sigma$, $\Xi$, $\rho$, $\omega$, and many others, thus creating a chaotic situation which inspired a lot of studies. In 1961, Murray Gell-Mann and Yuval Ne'eman proposed the eightfold way~\cite{GellMann:1961ky,Neeman:1961jhl} to classify all the then known hadronic states into SU(3)$_f$ multiplets and predicted the existence of $\Omega^-$, which was discovered in 1964~\cite{Barnes:1964pd}. In the same year, Gell-Mann~\cite{GellMann:1964nj} and Zweig~\cite{Zweig:1964jf} developed the now well-known quark model.   In the naive quark model, baryons are made of three quarks and mesons of a pair of quark and antiquark.  It has been very successful in explaining most if not all of the experimentally discovered baryons and mesons (see, e.g., Refs.~\cite{Godfrey:1985xj,Capstick:1986bm}).

In 2003, three new hadrons were discovered, i.e., $X(3872)$~\cite{Choi:2003ue}, $D_{s0}^*(2317)$~\cite{Aubert:2003fg}, and $\Theta^+(1540)$~\cite{Stepanyan:2003qr}, which opened a new era in hadron spectroscopy and our understanding of the low energy strong interaction. In the following years, many more exotic mesons were discovered, which do not fit easily into the conventional $qqq$ and $q\bar{q}$ picture and are collectively referred to $XYZ$ states~\cite{Brambilla:2019esw}. In 2015 and 2019, the LHCb Collaboration reported the discovery of four pentaquark states, $P_c(4380)$~\cite{Aaij:2015tga}, $P_c(4312)$, $P_c(4440)$, and $P_c(4457)$~\cite{Aaij:2019vzc}. Due to their masses and minimum quark contents $c\bar{c}qqq$, they are clearly beyond the $qqq$ baryon picture. In 2020, the LHCb Collaboration observed another structure in the $J/\psi\Lambda$ invariant mass distribution of the $\Xi_b^-\to J/\psi\Lambda K^-$ decay~\cite{Aaij:2020gdg}, which could be the strange counterpart~\cite{Liu:2020hcv,Lu:2021irg} of the $P_c$'s they discovered in 2015 and 2019. Although a large amount of theoretical and experimental studies have been performed to understand the nature of these states (for reviews, see, e.g., Refs.~\cite{Brambilla:2019esw,Liu:2013waa,Hosaka:2016pey,Chen:2016qju,Guo:2017jvc,Liu:2019zoy}), i.e., whether they are compact multiquark states or loosely bound states of more conventional hadrons or mixtures of conventional hadrons and multiquark states, we are still far away from a unified picture which can explain all their relevant properties.

One thing is clear, though. Because a majority of these exotic states are located close to the thresholds of two or more conventional hadrons, final state interactions can play an important role, to the extent that some of these states may qualify as hadronic molecules. Nonetheless, it is not an easy task to distinguish the molecular picture from its competing counterparts. The reason is pretty obvious. Quantum mechanically, a hadronic state can have all the configurations allowed by its quantum numbers. For instance, a meson can have both $q\bar{q}$ and $qq\bar{q}\bar{q}$ components. The latter can further be classified into a meson-meson configuration and the rest. We refer to a state as a conventional meson if the $q\bar{q}$ component is dominant, or a meson-meson molecule if the meson-meson configuration is dominant.  In the present work, we focus on the molecular picture.

To either confirm and refute the molecular picture, one can turn to nuclear physics. We know that atomic nuclei, to a large extent, can be treated as bound states of nucleons, i.e., a deuteron is composed of  one proton and one neutron, a triton is composed of one proton and two neutrons, a $^{12}$C nucleus contains six protons and six neutrons, and so on. Such a picture has been tested extensively. Nowadays, using nucleons as degrees of freedom and with the two-body $NN$ interaction determined by the nucleon-nucleon scattering data and the residual small $NNN$ interaction, ab initio calculations can reproduce most of the ground and low-lying excited states of light- and medium-mass nuclei. Adding hyperons to the nuclear system, one ends up with  hypernuclei. Following the same approach, the properties of hypernuclei can also be reproduced very well.  One must mention that nowadays the so-called nuclear ab initio calculations have already entered a precision era  (see the talk of Gaute Hagen in this proceeding).

Now turning  to exotic hadrons, if we believe that one of the exotic hadrons $C$ is a bound state of two other hadrons $A$ and $B$. Then using the same interaction between $A$ and $B$ which leads to the formation of $C$, one can study the three-body $ABB$ or $AAB$ system and check whether they are bound or not. If the $ABB$ or $AAB$ system binds, then an experimental confirmation on the existence of this state can unambiguously verify our molecular picture for the exotic state $C$, i.e., it is  dominantly a hadronic molecule.

In the present talk, we argue that the $D_{s0}^*(2317)$ state can be treated as a hadronic molecule of $DK$ and as a result $DDK$, $D\bar{D}K$, and $D\bar{D}^*K$ bind. An experimental discovery of any of these states can provide a highly nontrivial check on the molecular picture of $D_{s0}^*(2317)$ and of the many other exotic states discovered so far. To facilitate experimental searches for these states, we also use the effective Lagrangian approach to study their strong decays via triangle diagrams. Furthermore, we test our approach by studying two  well known systems, the triton and the $\bar{K}NN$ system.

\section{$D_{s0}^*(2317)$ as a $DK$ bound state}
\label{sec:1}
 $D_{s0}^*(2317)$ was first discovered by the BaBar Collaboration~\cite{Aubert:2003fg} and then confirmed by CLEO~\cite{Besson:2003cp} and Belle~\cite{Krokovny:2003zq} collaborations. It is located  45 MeV below the $DK$ threshold and has a decay width less than 3.8 MeV. The observed mass and width are far away from the predicted mass of 2460 MeV and width of hundreds of MeV in the quark model~\cite{Barnes:2003dj}. Thus  $D_{s0}^*(2317)$ is difficult to be interpreted as a conventional $c\bar{s}$ state. On the other hand, due to the strongly attractive $DK$ interaction, it can  be easily explained as a $DK$ molecule~\cite{Barnes:2003dj,Kolomeitsev:2003ac,Hofmann:2003je,Guo:2006fu,Gamermann:2006nm,Gamermann:2006nm,Liu:2012zya,Altenbuchinger:2013vwa,Mohler:2013rwa,Lang:2014yfa,Bali:2017pdv}.

 It is interesting to note that in the unitarized chiral approach the $D^*K$ interaction is the same as the $DK$ interaction up to heavy quark spin symmetry breaking effects. As a result, the existence of a $DK$ molecule implies the existence of a $D^*K$ molecule. In Ref.~\cite{Altenbuchinger:2013vwa}, fixing the next-to-leading order low-energy constants (LECs) and a subtraction constant by fitting to the lattice QCD  scattering lengths~\cite{Liu:2012zya}, and then solving the Bethe-Salpeter equation, one found two poles in the strangeness 1 and isospin 0 channel as shown in Table~\ref{DPeaks}, which coincide with the experimentally known $D_{s0}^*(2317)$ and $D_{s1}(2460)$. In such a picture, one can easily understand the fact that the mass difference between $D_{s1}(2460)$ and $D_{s0}^*(2317)$ is almost the same as the mass difference between $D^*$ and $D$ because in the molecule picture, the mass difference comes from the different masses of components, as the interactions between the components are the same due to heavy quark spin symmetry.  Furthermore, the UChPT also predicts the existence of two strangeness 0 and isospin 1/2 states. A recent analysis of the LHCb data seems to support the existence of the $0^+$ state~\cite{Du:2020pui}. If the $1^+$ state can also be discovered, it will provide a nontrivial check on the UChPT  and its explanation of $D_{s0}^*(2317)$ as a $DK$ bound state.  Using the heavy quark flavor symmetry, the UChPT  also predicts the existence of the bottom counterparts of $D_{s0}^*(2317)$, shown in Table~\ref{BPeaks}. Though they have not been discovered experimentally, the predicted masses are consistent with the lattice QCD simulations of Ref.~\cite{Lang:2015hza}.

\begin{table}[htpb]
      \renewcommand{\arraystretch}{1.2}
     \setlength{\tabcolsep}{0.1cm}
\caption{\label{DPeaks}Pole positions $\sqrt{s}=M-i\frac{\Gamma}{2}$ (in units of MeV) of charmed mesons dynamically generated in the  UChPT of Ref.~\cite{Altenbuchinger:2013vwa}. }
\centering
\begin{tabular}{c|c|c}
\hline\hline
  $(S,I)$ & $J^P=0^+$& $J^P=1^+$ \\
\hline
  (1,0) & $2317\pm10$ & $2457\pm17 $  \\
  (0,1/2)& $(2105\pm4)-i(103\pm7)$ & $(2248\pm6)-i (106\pm13)$\\
   \hline\hline
\end{tabular}
\end{table}

\begin{table}[htpb]
      \renewcommand{\arraystretch}{1.2}
     \setlength{\tabcolsep}{0.1cm}
\caption{\label{BPeaks}Pole positions $\sqrt{s}=M-i\frac{\Gamma}{2}$ (in units of MeV) of bottom mesons dynamically generated in the UChPT of Ref.~\cite{Altenbuchinger:2013vwa}.}
\centering
\begin{tabular}{c|c|c}
\hline\hline
  $(S,I)$ & $J^P=0^+$ & $J^P=1^+$  \\
\hline
  (1,0) & $5726\pm 28$ & $5778\pm26$  \\
  (0,1/2) & $(5537\pm14)-i(118\pm22)$ & $(5586\pm 16)-i(124\pm25)$ \\
   \hline\hline
\end{tabular}
\end{table}

\section{$DDK$, $D\bar{D}K$, and $D\bar{D}^*K$ bound states}
Now if we tentatively accept the picture where $D_{s0}^*(2317)$ is dominantly a $DK$ bound state, it is natural to ask what happens if we add one $D/\bar{D}/\bar{D}^*$ into the $DK$ pair. Will the resulting three-body systems bind? If they do, what are the binding energies and strong decay widths. Where can future experiments search for them?  In this section, we address these questions.

\subsection{Two-body interactions}
Inspired by the UChPT of Ref.~\cite{Altenbuchinger:2013vwa}, we parameterize the $DK$ interaction with two Gaussians, i.e.,
\begin{equation}
\label{eq:V2-final}
V_{WT}(\vec{r}) = C(R_c)\,e^{-(r/R_c)^2} + C_S\,e^{-(r/R_S)^2},
\end{equation}
where the first term represents attraction and the second term  possible short-range repulsion. $R_C$ and $R_S$ are coordinate space cutoffs representing the effective interaction ranges with the constraint of $R_s<R_c$. The $\bar{D}K$ interaction is taken to be half that of the $DK$ interaction according to Ref.~\cite{Altenbuchinger:2013vwa}, while the $\bar{D}^*K$ interaction is the same as the $\bar{D}K$ interaction because of heavy quark spin symmetry. For the $DD$ and $D\bar{D}^{(*)}$ interactions, we turn to the one-boson exchange (OBE) model. The explicit form for the $DD$ potential can be found in Ref.~\cite{Wu:2019vsy}, from which one can obtain the $D\bar{D}$ potential, while  the $D\bar{D}^*$ potential can be found in Ref.~\cite{Liu:2019stu}, where the cutoff of the regulator is fixed by reproducing $X(3872)$ as a bound state of $\bar{D}D^*$. The same cutoff is used for the $D\bar{D}$ interaction. We note that this cutoff does not yield a $D\bar{D}$ bound state as claimed by the lattice QCD simulation of Ref.~\cite{Prelovsek:2020eiw}. A larger cutoff is needed for such a scenario, which will increase the binding energy of the $D\bar{D}K$ state by about 20 MeV, but does not change qualitatively our following results.
\subsection{Binding energies and radii}
\begin{figure}[!h]
  \centering
 \includegraphics[scale=0.5]{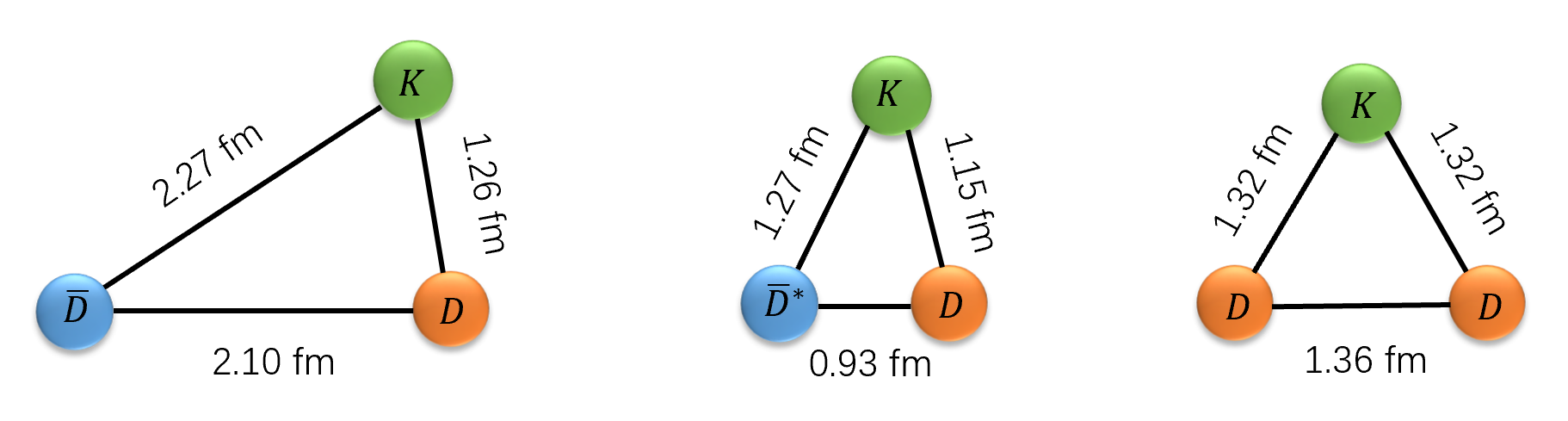}
  \caption{ RMS radii of subsystems in the $D\bar{D}K$ (left), $D\bar{D}^*K$ (middle), and $DDK$~\cite{Wu:2019vsy}  bound states with a cutoff $R_c=1.0$ fm. Taken from Ref.~\cite{Wu:2020job}.}
  \label{RMS}
\end{figure}

Now we present some numerical results. The binding energies of the $DDK$, $D\bar{D}K$, and $D\bar{D}^*K$ bound states are given in Table~\ref{Results:BE}, in comparison with the results of other works. Among the three bound states, the $D\bar{D}^*K$ state has the largest binding energy, because the $D\bar{D}^*$ interaction is attractive enough to dynamically generate $X(3872)$ and $Z_c(3900)$. The uncertainties are due to  the consideration of the likely existence of a short rang repulsion and the use of several cutoffs (see Refs.~\cite{Wu:2019vsy,Wu:2020job} for more details). In Fig.~\ref{RMS}, we show the root mean square (RMS) radii of the two-body subsystems of the three bound states.  Consistent with the binding energies shown in Table~\ref{Results:BE}, the $D\bar{D}K$ system is more extended, while the $D\bar{D}^*K$ state is more compact. One should note that the spatial distributions are more sensitive to the details of the two-body potentials and the results shown in Fig.~\ref{RMS} can only serve for illustration purposes.
\begin{table}[htpb]
    \caption{Binding energies (in units of MeV) of the $DDK$, $D\bar{D}K$, and $D\bar{D}^*K$ bound states, in comparison with the results of other works, if available.\label{3b:BE}}
    \centering
    \begin{tabular}{c|c c c}
    \hline
    \hline
    System & \multicolumn{3}{c}{Binding energy}\\
    \hline
    $\frac{1}{2}(0^-)$ $DDK$ &  $67.1\sim71.2$~\cite{Wu:2019vsy}& $50\sim70$~\cite{SanchezSanchez:2017xtl} & $-$\\
       $\frac{1}{2}(0^-)$ $D\bar{D}K$&$48.9^{+1.4}_{-2.4}$~\cite{Wu:2020job}  & $-$ &$-$\\
       $\frac{1}{2}(1^-)$ $D\bar{D}^*K$&$77.3^{+3.1}_{-6.6}$~\cite{Wu:2020job}&$53.52^{+6.55}_{-6.13}$~\cite{Ma:2017ery} & $64\pm2$~\cite{Ren:2018pcd}\\
    \hline\hline
    \end{tabular}
    \label{Results:BE}
\end{table}

\subsection{Strong decays}
\label{sec:2}
\begin{figure}[htbp]
\begin{center}
\includegraphics[scale=0.45]{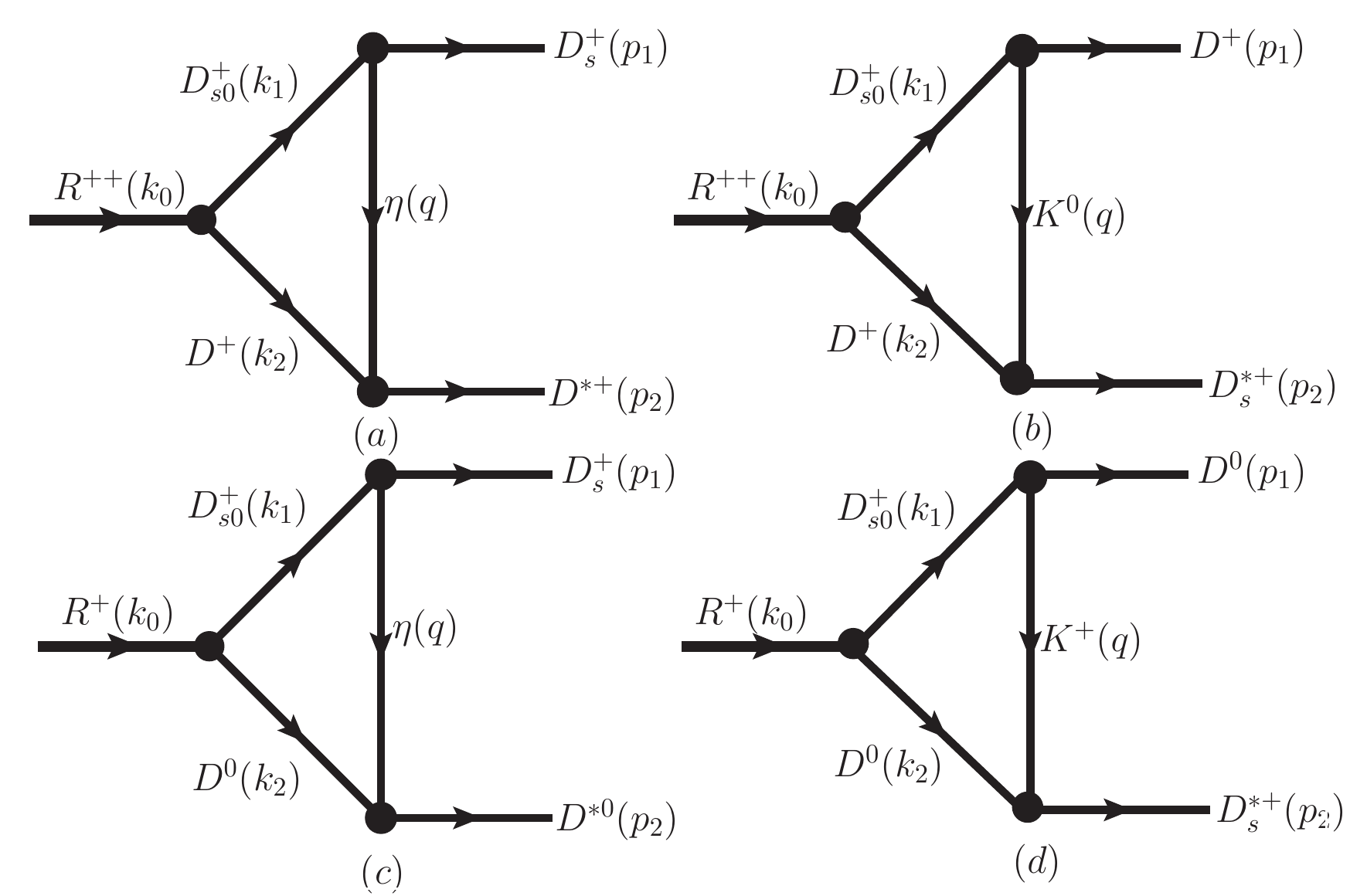}
\caption{Strong decay of the $R^{++}$ state to $D_s^{+}D^{*+}$ and $D^{+}D_s^{*+}$(a-b), and $R^{+}$ state to $D_s^{+}D^{*0}$ and $D^{0}D_s^{*+}$(c-d). Taken from Ref.~\cite{Huang:2019qmw}.}\label{mku}
\end{center}
\end{figure}

\begin{figure}[!h]
\begin{center}
\begin{tabular}{cc}
\begin{minipage}[t]{0.4\linewidth}
\begin{center}
\begin{overpic}[scale=.6]{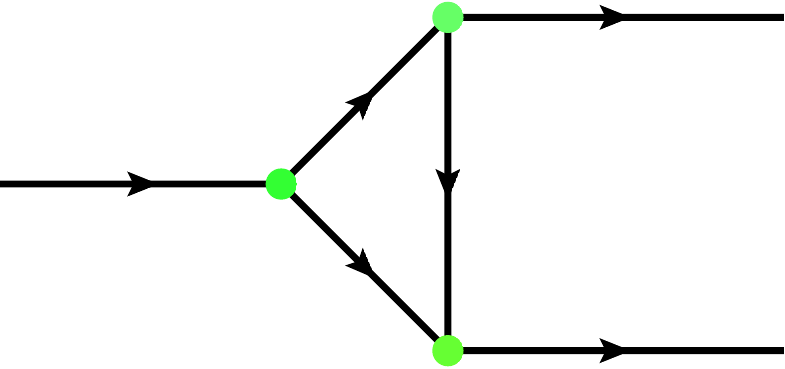}
		\put(74,6){$J/\psi$}
		
		\put(37,9){$\bar{D}$}
		
		\put(37,38){$D_{s0}^*$}
		
		\put(16,26){$K_{c}$ }
		\put(75,38){$K$} \put(60,22){$D$}
\end{overpic}
\end{center}
\end{minipage}
&
\begin{minipage}[t]{0.4\linewidth}
\begin{center}
\begin{overpic}[scale=.6]{triangle.pdf}
		\put(74,6){$\bar{D}^{\ast}$}
		
		\put(37,9){$\bar{D}$}
		
		\put(37,38){$D_{s0}^*$}
		
		\put(16,26){$K_{c}$ }
		\put(75,38){$D_{s}$} \put(60,22){$\eta$}
\end{overpic}
\end{center}
\end{minipage}
\end{tabular} \nonumber
\caption{Strong decays of $K_{c}(4180)$  to $J/\psi K$ and $D_{s}\bar{D}^{\ast}$ via triangle diagrams. Taken from Ref.~\cite{Wu:2020job}.\label{decay}}
\end{center}
\end{figure}
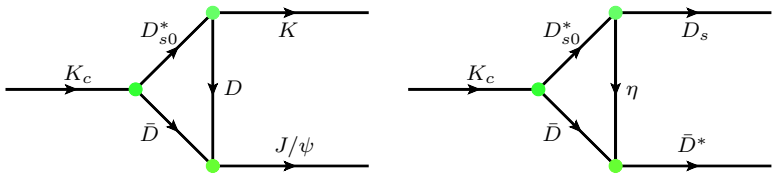

In order to know where to search for these states, we have studied the strong decays of $DDK$ and $D\bar{D}K$ via triangle diagrams as shown in Fig.\ref{mku} and Fig.~\ref{decay}.~\footnote{For the two-body decay of the
$D\bar{D}^*K$ bound state, see Ref.~\cite{Ren:2019umd}.} We will not go into the numerical details, which can be found in Ref.~\cite{Huang:2019qmw} and Ref.~\cite{Wu:2020job}, respectively. It is enough to note that the total decay width of the $DDK$ bound state is about $2\sim3$ MeV, while that of the $D\bar{D}K$ bound state is less than 1 MeV.

The Belle Collaboration has searched for the $DDK$ bound state in  $\Upsilon(1S,2S)$ inclusive decays and via direct production in $e^+e^-$ collisions at $\sqrt{s} = 10.520$, $10.580$, and $10.867$ GeV and no significant signals are observed in the $D^+ D_{s}^{*+}$
 invariant-mass spectra of all the modes studied~\cite{Li:2020gvn}. It is not surprising because due to the doubly charmed nature of the $DDK$ bound state, its production yield in $e^+e^-$ collisions might be small.  Belle II will collect 50 times more data and may have a better chance to observe it.

 On the other hand, the $D\bar{D}K$ state can be observed either via the $J/\psi K$ final state or the $D_s\bar{D}^*$ final state. Luckily, LHCb~\cite{Aaij:2015tga,Aaij:2017zgz}, Belle~\cite{Shen:2014gdm}, and BESIII~\cite{Ablikim:2020hsk} all have access to these final states. We can only encourage these experiments to search for it in the future. It should be mentioned  that the recently observed $Z_{cs}(4220)$ by the LHCb Collaboration~\cite{Aaij:2021ivw} though has a mass close to the $D\bar{D}K$ state, but the quantum numbers are different. The preferred spin-parity for the $Z_{cs}(4220)$ state is $1^+$, while the $D\bar{D}K$ state has spin-parity $0^-$.

\section{From deuteron to triton}
It is instructive to show that the extension from a two-body system to the related three-body system is relatively reliable. One such system is the deuteron and the triton. For such a purpose, we use two types of $NN$ potentials to study the $NN$ and $NNN$ systems.
For the first case, we follow the approach of studying the $DDK$ system and use the Gaussian shape potential in Eq.~(\ref{eq:V2-final}) with $C_S=0$ (hence only $S$-wave interaction is considered). We choose the cutoff $R_c$ and interaction strength $C(R_c)$ to reproduce the binding energy 2.22 MeV and RMS radius 4.0 fm of deuteron (with isospin $I=0$). The isospin $I=1$ $NN$ interaction is known to be attractive but cannot form a bound state. The maximum strength of the interaction can be obtained by gradually decreasing the strength of the  $I=1$ interaction from that of the $I=0$ interaction until the $I=1$ system becomes unbound, yielding a critical value for $C(R_c)$. For  the second case, we use the $NN$ OBE potential with a dipole form factor, for which we choose the cutoff $\Lambda=0.859$ GeV to reproduce the binding energy of deuteron. For more details about the OBE potential, see Ref.~\cite{Wu:2020rdg}

With the $NN$ potentials mentioned above, we study the $NNN$ system with the Gaussian expansion method, the results are presented in Table~\ref{DeuToTri}. Clearly, the resulting binding energy of $8.16\sim5.6$ MeV is consistent with the experimental binding energy of 8.48 MeV and the RMS radii from the two potentials are also relatively close.  This naive exercise demonstrated the reliability of the approach we adopted.

\begin{table}[htpb]
\begin{center}
\caption{Binding energies (in units of MeV) and RMS radii (in units of fm) of deuteron and triton. }
\label{DeuToTri}
\begin{tabular}{ c c c c c}
\hline\hline
 & \multicolumn{2}{c}{Gaussian} & \multicolumn{2}{c}{OBE}  \\
 Systems & B.E. & $\sqrt{\langle r^2\rangle}$& B.E. & $\sqrt{\langle r^2\rangle}$ \\
 \hline
Deuteron & 2.22 & 4.02& 2.22 & 3.82 \\
Triton & 8.16 & 3.42& 5.60 & 4.03\\ \hline\hline
\end{tabular}
\end{center}
\end{table}

\section{From $\bar{K}N$ to $\bar{K}NN$}
Another system that resembles the $DDK/D\bar{D}K$ system of our main interest is the $\bar{K}NN$ system,  because the $\Lambda(1405)$ state has long been described as a $\bar{K}N$ molecule and the strong $\bar{K}N$ interaction also leads to the formation of $\bar{K}NN$ bound states,
as has been extensively studied both theoretically~\cite{Yamazaki:2002uh,Magas:2006fn,Shevchenko:2006xy,Ikeda:2007nz,Arai:2007qj,Nishikawa:2007ex,Dote:2008hw,Wycech:2008wf,Uchino:2011jt,Barnea:2012qa}
and experimentally~\cite{Agnello:2005qj,Yamazaki:2008hm,Yamazaki:2010mu,Tokiyasu:2013mwa,Ichikawa:2014ydh,Ajimura:2018iyx,Sada:2016nkb},
with the later supporting the existence of the $\bar{K}NN$ state. It is interesting to check whether following the same approach we used to study the $DDK$ and $D\bar{D}K$ systems, we could obtain the $\bar{K}NN$ states, which also serve as a nontrivial check on our approach.

The $\bar{K}N$ interaction is the same WT term
which is responsible for the binding of the $D K$  system.
Using the WT term or the chiral potential as kernel to the Lippmann-Schwinger
or Bethe-Salpeter equations, one can describe the $\Lambda(1405)$ state as
a $\bar{K}N$ bound state~\cite{Oset:1997it,Kaiser:1995eg,Oller:2000fj,Lutz:2001yb,Jido:2003cb,Borasoy:2004kk,Hyodo:2007jq}.
For fixing the strength of the $\bar{K}N$ interaction we simply reproduce
the binding energy of the $\Lambda(1405)$ with the potential of
Eq.~(\ref{eq:V2-final}).
For the $NN$ interaction we use the OBE model.
The binding energy of the $\bar{K}NN$ state is listed
in Table \ref{Results:NNKbar} for different cutoffs.
In the same table we also show the values of the coupling that reproduces
the $\Lambda(1405)$ pole as a $\bar{K}N$ bound state with a binding energy
$B_2 = 29.4\,{\rm MeV}$.
The binding energy of the $\bar{K}NN$ state ranges from $35-42\,{\rm MeV}$,
where the cutoff dependence is relatively weak.
One should note that the study of the $\bar{K}NN$ system presented here should be viewed mostly as a consistency check on the approach proposed because there is a large
literature on the $\bar{K}NN$ system with more sophisticated formalisms.  Nevertheless, both the predicted binding energy and RMS radii are qualitatively  consistent with those of more sophisticated models, such as that of Ref.~\cite{Dote:2008hw}, which predicts a binding energy of $12\sim22$ MeV, a RMS radius of $r_3(NN)$ of $2.09\sim2.26$ fm and $r_3(\bar{K}N)$ of $1.85\sim2.01$ fm.

\begin{table}[htpb]
	\centering
	\caption{Binding energies ($B_2$ and $B_3$ in MeV) and RMS radii (in units of MeV) of the $\bar{K}N$ and $I(J^P)$=$\frac{1}{2}(0^-)$ $\bar{K}NN$ bound states. The parameters are determined by reproducing the $\Lambda(1405)$ with a binding energy 29.4 MeV with respect to the $\bar{K}N$ threshold.}\label{Results:NNKbar}
	\begin{tabular}{ c c c c c c c}
		\hline
		\hline
		$C(R_c)$& $R_c$ & $B_2(\bar{K}N)$&$r_2(\bar{K}N)$ & $B_3(\bar{K}NN)$&$r_3(NN)$&$r_3(\bar{K}N)$\\
		\hline
		&&$C_S=0$& &$R_s=0.1$& &\\
		\hline
		$-925.9$ & $0.5$ & $29.4$ &1.28 & $35.2$&2.07 &2.64 \\
		$-316.4$ & $1.0$ & $29.4$ & 1.55& $39.3$&1.99 &2.39\\
		$-132.6$ & $2.0$ & $29.4$ &2.05 & $41.8$&2.34 &2.69 \\
		\hline
		&&$C_S=1000$& &$R_s=0.1$&&\\
		\hline
		$-946.6$ & $0.5$ & $29.4$ &1.28 & $35.4$&2.06 &2.63 \\
		$-319.8$ & $1.0$ & $29.4$&1.55 & $39.4$& 1.99&2.39 \\
		$-133.2$ & $2.0$ & $29.4$&2.05 & $41.8$&2.34 &2.69 \\
		\hline
		\hline
	\end{tabular}
\end{table}

\section{Summary and outlook}
We have argued that one can verify the molecular picture of $D_{s0}^*(2317)$ by studying the $DDK$, $D\bar{D}K$, and $D\bar{D}^*K$ three-body systems. We showed that indeed these systems bind. We also computed their strong decays via triangle diagrams. We encourage future experimental searches for these states. If discovered, they will not only decisively advance our understanding of the so-called exotic states but also imply the existence of new kinds of nuclear chart or periodic table.

\begin{acknowledgements}
We thank M. P. Valderrama, E. Hiyama, Alberto Martínez Torres, Kanchan P. Khemchandani, Xiu-Lei Ren,  Yin Huang, and  Ya-Wen Pan for collaborations on some of the topics covered in this talk.
\end{acknowledgements}

\bibliographystyle{spphys}       
\bibliography{multihadron}   

\end{document}